\documentclass[lettersize,journal]{IEEEtran}
\usepackage{amsmath,amsfonts}
\usepackage{algorithmic}
\usepackage{algorithm}
\usepackage{array}
\usepackage[caption=false,font=normalsize,labelfont=sf,textfont=sf]{subfig}
\usepackage{textcomp}
\usepackage{stfloats}
\usepackage{url}
\usepackage{verbatim}
\usepackage{graphicx}
\usepackage{cite}
\begin{document}
\title{Threat Modelling and Risk Analysis for Large Language Model (LLM)-Powered Applications}
\author{Stephen Burabari Tete, ~\IEEEmembership{Teesside University UK}}
 
\markboth{Teesside University, School of Computing, Engineering \& Digital Technology, United Kingdom   - January ~2024}
{Shell \MakeLowercase{\textit{et al.}}: Teesside University for IEEE Journals}

\maketitle

\begin{abstract}
The advent of Large Language Models (LLMs) has revolutionized various applications by providing advanced natural language processing capabilities. However, this innovation introduces new cybersecurity challenges. This paper explores the threat modeling and risk analysis specifically tailored for LLM-powered applications. Focusing on potential attacks like data poisoning, prompt injection, SQL injection, jailbreaking, and compositional injection, we assess their impact on security and propose mitigation strategies. We introduce a framework combining STRIDE and DREAD methodologies for proactive threat identification and risk assessment. Furthermore, we examine the feasibility of an end-to-end threat model through a case study of a custom-built LLM-powered application. This model follows Shostack's Four Question Framework, adjusted for the unique threats LLMs present. Our goal is to propose measures that enhance the security of these powerful AI tools, thwarting attacks, and ensuring the reliability and integrity of LLM-integrated systems.
\end{abstract}

\begin{IEEEkeywords}
cybersecurity, threat modelling, risk analysis, large language model, LLM, AI.
\end{IEEEkeywords}

\section{Introduction}

\IEEEPARstart{A}{s} the technological domain continues to embrace the power of artificial intelligence, Large Language Models (LLMs) are at the forefront of this revolution, transforming how we interact with and benefit from machine intelligence. LLMs have demonstrated an uncanny ability to understand and generate human-like text, propelling a wide array of applications to new heights of capability. With this evolution, identifying and mitigating the underlying risks through robust threat modeling and risk analysis is becoming a critical necessity for maintaining the integrity and trustworthiness of LLM-powered systems. Groundbreaking studies and meticulous surveys have laid the foundation for understanding the risks associated with these models, highlighting the dual nature of technology as both an asset and a vulnerability \cite{bingsearch2023}.

\subsection{LLM-Powered Applications}
LLMs such as GPT-4 and PaLM 2 signify a leap forward in natural language processing, boasting advanced capabilities in understanding context and producing relevant responses \cite{brown2020language,anil2023palm2}. These models have become integral in numerous real-life applications, ranging from augmented search engines to sophisticated virtual assistants. For instance, GPT-4's integration into Microsoft's Bing Search is a testament to its practical utility, while Google's Bard demonstrates the potential of PaLM 2 as a powerful information retrieval tool \cite{bingsearch2023,pichai2023importantstepai}.

Beyond the realm of search and retrieval, LLM-integrated applications are redefining boundaries in other sectors. In healthcare, LLMs are exploring uncharted territories, from enhancing patient communication through AI chatbots to long-term risk prediction for chronic conditions, offering invaluable insights into preemptive care \cite{chatgptplugins2023,chatwithpdf2023,introducingchatgpt2023,jailbreakinggpt42023,learnprompting2023,llama213bchaturl2023,llama27bchaturl2023}. Each application harnesses the strengths of these models, from analyzing prompts to synthesizing responses in intricate tasks like spam detection, translation, and more. By combining instruction prompts guided by human intent with data prompts sourced from external resources, these applications provide tailored and contextually apt solutions for users \cite{brown2020language,liu2022goodincontext,shin2022pretraining,wei2022chaintought}.

\subsection{The Threats and Risks}

Harnessing the transformative power of LLMs does not come without its threats and risks. As the applications grow more complex and integral to our infrastructure, the potential repercussions of security breaches or faulty outputs become increasingly severe. Gleaning insights from comprehensive surveys on machine-generated text, we are reminded that threat models should be meticulously crafted to detect, analyze, and mitigate any form of malicious infiltration or unintended consequence \cite{bingsearch2023}.

The integration of LLMs in sensitive sectors like healthcare further underscores the urgency for robust threat modeling. With studies illustrating the promise of LLMs in managing health-related risks, it is evident that ensuring their security is non-negotiable \cite{chatgptplugins2023,chatwithpdf2023,introducingchatgpt2023,jailbreakinggpt42023,learnprompting2023,llama213bchaturl2023,llama27bchaturl2023}. Keeping confidential patient data safe while providing accurate predictions and diagnostics requires an in-depth understanding of cyber threats, ranging from data breaches to sophisticated adversarial attacks that could exploit model vulnerabilities.

The management of these threats begins with identifying the scope of potential risks, guided by frameworks that consider the entire lifecycle of the system \cite{bingsearch2023}. From development through to deployment, each phase presents unique challenges that must be addressed proactively. Through the adoption of security development life cycles and consistent security reviews, as Microsoft has demonstrated with the trustworthy computing security development lifecycle, we can significantly reduce the rate of vulnerabilities in software underpinned by LLMs \cite{brown2020language,pichai2023importantstepai}.

In this paper, our primary goal is to examine the threats and risk of LLM-powered applications, as well as possible mitigation to prevent or reduce the threats and risks. In particular, we focus on the following research questions (RQ):
\begin{itemize}
\item{\textbf{RQ1: What are the potential attacks against LLM-powered applications and what are their impact on the application's security?} To study this question, we focused on web applications built upon Flask framework, using OpenAPI's GPT-3.5 and GPT-4 models.}
\item{\textbf{RQ2: What defence or mitigation can prevent or reduce the effectiveness of these attacks?} To answer this question, we reviewed previous mitigation provided and validated by several scholars and featured it together.}
\item{\textbf{RQ3: Is an end-to-end threat model for LLM-powered application feasible?} To address this question, we took a case study of our custom built LLM-Doctor \cite{tete_llm_doctor} application, powered by GPT-3.5 and GPT-4 and designed a custom end-to-end threat model.}
\end{itemize}

\section{Overview of Threat Modelling, and \\ Risk Analysis}
\subsection{Threat Modelling (TM)}

\begin{figure}
    \centering
    \includegraphics[width=1\linewidth]{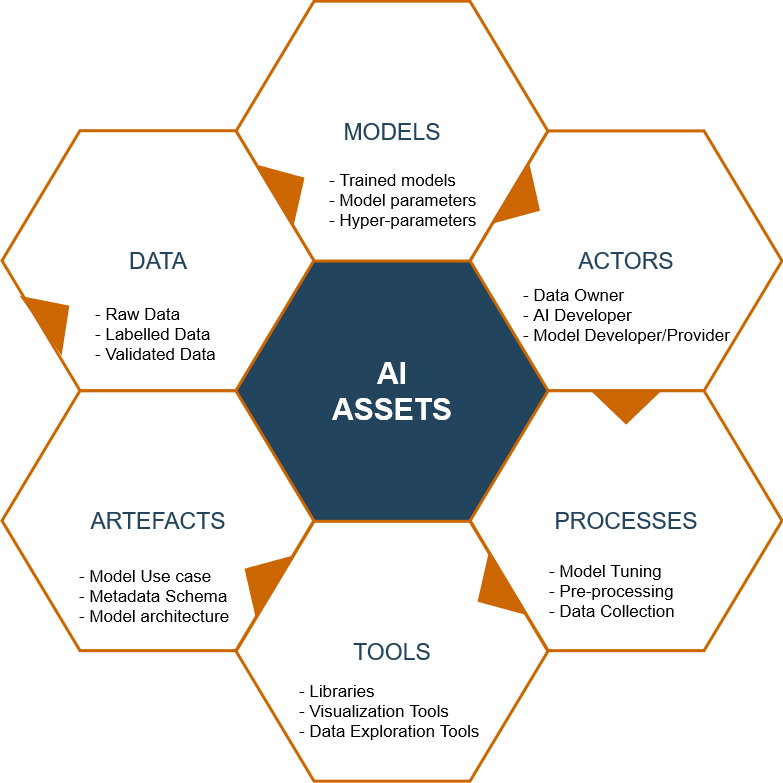}
    \caption{AI Assets}
    \label{fig:enter-label}
\end{figure}
Threat modelling is a structured approach for identifying and prioritizing potential threats to a system and determining the value that potential mitigations would have in reducing or neutralizing those threats. This process is essential in the proactive identification and management of cybersecurity risks, particularly in complex distributed systems, such as those that harness the capabilities of Large Language Models (LLMs).

The evolution of threat modelling has paralleled advancements in the software development life cycle (SDLC), with various methodologies being developed to suit distinct environments. Initially conceptualized through the CIA triad proposed by the Johnson Space Center in the 1980s, which underlined the importance of confidentiality, integrity, and availability, threat modeling methodologies have expanded to include additional elements such as authentication, non-repudiation, and authorization, as seen in the STRIDE model \cite{kim2022stride, threat-modelling-review}.

One of the seminal works in threat modelling was introduced by Schneider in 1998 with the use of attack trees as a means of graphically representing and analyzing security threats, providing a foundational approach to threat modeling \cite{adapting-threat-modelling}. Today, threat modeling techniques such as STRIDE (an acronym for Spoofing, Tampering, Repudiation, Information Disclosure, Denial of Service, and Elevation of Privilege) provide a systematic and comprehensive method for identifying potential security threats to software applications \cite{threat-modelling-aiml}).

Moreover, the demand for rigorous security measures throughout the SDLC is apparent in the Trustworthy Computing Security Development Lifecycle (SDL) advanced by Microsoft \cite{lipner2004trustworthy}, in which software undergoes a final security review by an independent team and demonstrated a significant reduction in the discovery of security vulnerabilities compared to non-SDL-adhered software.

As machine learning (ML) and artificial intelligence (AI) become increasingly integral to numerous applications, the need to model threats specifically to AI-ML systems is more critical than ever \cite{dietterich2017steps}; \cite{threat-modelling-aiml}). This is particularly important as we continue moving towards systems with substantial autonomy, requiring a comprehensive life cycle approach to the design and development of AI solutions \cite{mireshghallah2023llms}.

\subsection{TM Framework}
Given the complexity and novelty of LLM-powered applications, it is of paramount importance to address their security comprehensively. Our proposed framework for threat modeling and risk analysis combines strategies from established methodologies that address both the intricacies of AI systems and existing best practices in security assessment.

STRIDE, with its categorization of threats, serves as the ideal foundational model for threat modeling in LLM-powered applications. It comprehensively covers the multitude of potential vulnerabilities and the rising need to fortify AI-ML systems against increasingly sophisticated threats \cite{threat-modelling-aiml}). The inclusion of elements from Shostack's Four Question Framework enhances the effectiveness of STRIDE, facilitating an improved understanding of security goals, potential attackers, likely attack vectors, and the impact of successful attacks.

\subsection{Risk Analysis}
DREAD, an acronym for Damage, Reproducibility, Exploitability, Affected Users, and Discoverability, offers a risk analysis methodology that allows for a more granular evaluation of the severity of potential threats identified during the threat modelling process. The coupling of STRIDE with DREAD aligns with the enhancement of threat modeling outcomes, as evidenced by a study focused on distributed control systems in the oil industry \cite{kim2022stride}).

Furthermore, the complexity and advanced capabilities of LLMs, evidenced by state-of-the-art natural language generation (NLG) systems, necessitates a threat modeling framework that accounts for new vectors of abuse specific to such systems \cite{machine_generated_tx}). Therefore, our proposed framework not only incorporates classic methodologies but also addresses the novel risks posed by the deployment of LLM-powered applications.

Our comprehensive review of cybersecurity risk assessment methods for complex systems underscores the importance of strategic planning throughout the system lifecycle, which is palpable in the integration of threat modeling and risk analysis from the conception phase of LLM-powered applications through to their production and deployment \cite{threat-modelling-review}).

By integrating STRIDE and Shostack's Four Question Framework for effective threat identification, and employing DREAD for in-depth risk analysis, the proposed framework intends to elevate the security posture of LLM-powered applications throughout their entire life cycle, thus ensuring robustness against an evolving threat landscape.

\section{Potential Attacks on LLM-powered Apps (RQ1)}
\begin{figure}
    \centering
    \includegraphics[width=1.0\linewidth]{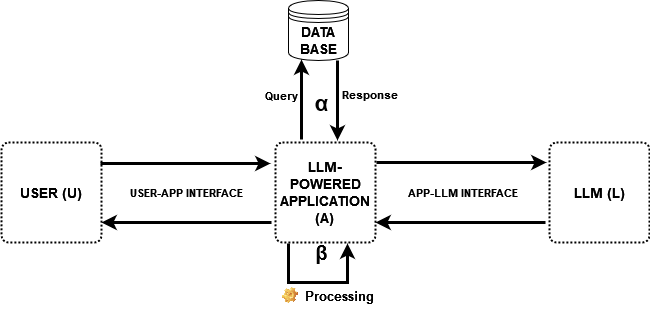}
    \caption{LLM-Powered application scheme }
    \label{fig:enter-label}
\end{figure}
With the increasing integration of Large Language Models (LLMs) across various domains, an understanding of potential security threats has become paramount. Given their deep interconnection with data and subjective decision-making capabilities, LLM-powered applications are susceptible to a myriad of sophisticated attacks. This section explores a range of potential attacks, leveraging various research works to delineate the nature and feasibility of these adversarial strategies.

\subsection{Data Poisoning Attack}
\begin{figure}
    \centering
    \includegraphics[width=1\linewidth]{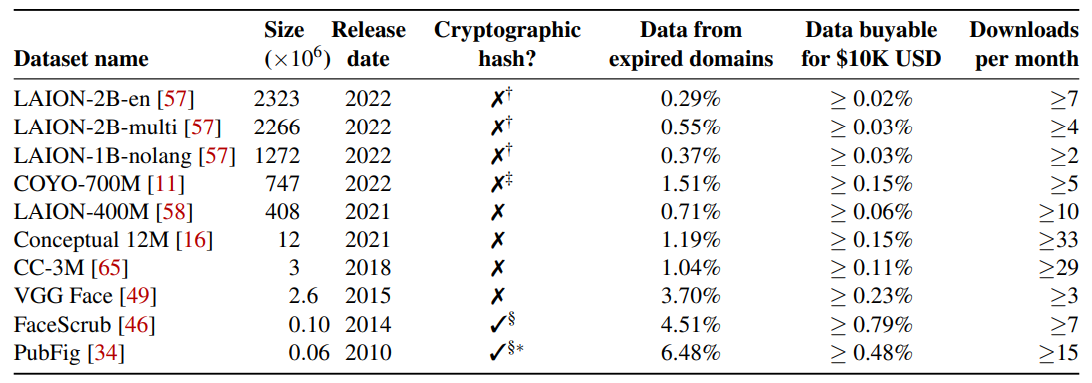}
    \caption{All recently-published large datasets are vulnerable to split-view poisoning attacks.  (Carlini et al., 2023)}
    \label{fig:enter-label}
\end{figure}
Data poisoning is a pernicious form of attack employing the subtle manipulation of training data. By contaminating the dataset, an adversary can skew the model's behavioral patterns, potentially causing intended misclassification or biased outputs. Carlini et al. (2023) highlight the significant risks associated with training deep learning models on large-scale datasets amassed from the internet \cite{carlini2023poisoning}. These datasets are vulnerable because they rely on the mutable nature of web content, which can be manipulated to include poisoning instances intended to skew the model's learned behavior. The research emphasizes how poisoning can be both straightforward and cost-effective, making it a tangible threat to the integrity of machine learning models \cite{Poisoning-Web-Scale-Training-Datasets2023}.

\textbf{Techniques for Data Poisoning}
\begin{enumerate}
    \item Split-View Poisoning
   This method takes advantage of the dynamic nature of internet resources. By altering the content after it has been initially indexed, attackers can ensure that what gets included in the dataset later differs from the original. Such discrepancies can mislead annotators and poison the dataset, influencing the bias and the decisions of the trained models\cite{Poisoning-Web-Scale-Training-Datasets2023}.
\item Frontrunning Poisoning
   Data poisoning can also occur in datasets that snapshot user-generated or crowd-sourced content at regular intervals. By timing malicious changes to appear during these snapshots, attackers can insert poisoning examples without needing to maintain long-term control over the content\cite{Poisoning-Web-Scale-Training-Datasets2023}.
\end{enumerate}

\textbf{Implications of Data Poisoning:}
These poisoning techniques undermine trust in the data's integrity and raise concerns about the safety and fairness of the resulting models. An adversary can strategically introduce errors leading to selected misclassification or biased model outputs. The ease and affordability of these attacks underscore the urgency to develop robust mitigation strategies.

\subsection{Prompt Injection Attacks}

A quintessential vulnerability in LLMs is the exploitation of the prompt mechanism. Attackers craft malicious inputs designed to deceive or misguide the model into producing unintended outcomes. Yi et al. (2023) and Liu et al. (2023) emphasize the severity of this issue, illustrating how indirect prompt injection attacks can present significant challenges to the integrity of LLMs \cite{promp_injection2023benchmarking}, \cite{prompt_injection_liu2023prompt}. This vulnerability can cascade severely, affecting LLM-integrated web applications, such as Rodrigo Pedro and colleagues have shown in their work, potentially leading to SQL injection attacks, pivoting from seemingly innocuous prompt manipulations~\cite{sql_injection_2023prompt}.

\subsection{SQL Injection Attacks}
The long-standing plague of SQL injection continues to evolve with the advent of LLMs. Attackers leverage crafted prompts or inputs to coerce an LLM into generating or executing harmful SQL code. This type of intrusion can jeopardize data integrity as well as confidentiality. The work by Kindy and Pathan (2013), although predating current LLM complexities, still provides foundational insights into understanding such invasive techniques \cite{sql_injection_n_remedies_2013detailed}. Furthermore, Jahanshahi et al. (2020) discuss mitigation strategies, indicating that the robustness against SQL injection attacks remains an ongoing concern and necessitates novel defence mechanisms tailored to LLM environments \cite{Jahanshahi_2020}.

\subsection{Jailbreaking Attacks}
Jailbreaking \cite{jailbreak_li2023multistep} attacks directly confront the restrictive measures placed within LLMs, effectively 'breaking out' of the intended operational parameters. Mehrotra et al. (2023) and Wu et al. (2023) both address the alarming efficiency with which attackers can execute self-adversarial strategies to jailbreak black-box models, including highly advanced versions like GPT-4 \cite{jailbreaking_llm_mehrotra2023tree}\cite{wu2023_jailbreaking}. Zhang et al. (2023) and Robey et al. (2023) explore defense mechanisms, emphasizing the need for adaptive and resilient model structures to mitigate such risks \cite{zhang2023_jailbreaking_llm_defending}\cite{robey2023_defining_llm_jailbreaking}.

\subsection{Compositional Injection Attacks}
This attack vector introduces a complex combination of instructions that embed hidden attacks within an otherwise benign prompt. Jiang et al. (2023) present the 'Prompt Packer' concept, where an adversary deceives an LLM through multi-layered, compositional instructions with nefarious intent \cite{compositional_injection_jiang2023prompt}. Such compositional injections could bypass conventional safeguard measures and achieve an unauthorized influence on model behavior.

\subsection{Insecure Output Handling}
As the number two (2) vulnerability on the OWASP 2023 Top 10 for LLM application, insecure output from LLM can lead to: CRSF, SSRF, and XSS, if the output is exploited successfully. \cite{owasp2023} 

\textbf{Attack scenario:}
An LLM-powered application processes output from the LLM directly, without first validating the output. This enables the attacker to manipulate the system, by executing harmful output from the LLM, which the attacker delibrately caused through crafted prompt.

\section{Mitigation to attacks on LLM-powered Apps (RQ2)}
\subsection{Mitigation for Data Poisoning}
\textbf{Mitigating Frontrunning Poisoning}
\textbf{ Randomized Timing:}
By randomizing when snapshots of data are taken or by extending the time required to snapshot an entire dataset curators make it more difficult for an attacker to predict when to execute the poisoning.

\textbf{Review and Freeze Period: }
Instituting a review period where edits to the content are frozen before officially capturing the snapshot allows for the detection and removal of any malicious modifications.

\textbf{Mitigating Poisoning in General}
\textbf{Relying on Consensus:}
   For even broader datasets like Common Crawl, where poisoning prevention is more challenging, one proposed solution is to employ a consensus-based approach. Trust would be conferred on data elements that appear consistently across a range of sources, thus requiring attackers to compromise many instances, paralleling distributed systems' trust mechanisms.
\textbf{Developing Unique Defenses}
   Given the variety of data use cases and the complexities involved in understanding how content is used and interpreted by learning algorithms, specific solutions are required. These defenses would need to rely on in-depth knowledge of the data consumption and training process, and would vary based on the dataset and potential attack vectors.

While the \cite{liu2023prompt} outlines potential mitigation strategies, the authors also acknowledge the need for further research on application-specific solutions. By understanding the complexities of content vectorization and conflict resolution in the training process, more effective defenses against data poisoning could be developed.

\subsection{Mitigation for prompt injection}
Prompt injection \cite{liu2023-Prompt-Injection-Attacks}, \cite{promp_injection2023benchmarking} comes with different attack techniques, style and pattern.
Indirect, direct injection.
Prompt injection attacks present a significant threat to the trustworthiness and security of Language Model (LM) integrated applications. These attacks exploit weaknesses in the processing of prompts by the underlying models, potentially leading to incorrect, unwanted, or even malicious responses. In this section, we explore various mitigation strategies, with a focus on both prevention-based and detection-based defenses, to safeguard these systems.

\textbf{Prevention-Based Defenses:}
Prevention is the first line of defense against prompt injection attacks, aiming to safeguard the system before any damage can occur.

\begin{enumerate}
    \item Paraphrasing:
    Paraphrasing disrupts the structure of injection attacks by altering the sequence of characters and words, making it difficult for injected instructions to remain coherent within the altered data prompt. This technique was initially developed to defend against adversarial prompts \cite{zou2023universal} but has since been adapted to address prompt injection threats \cite{liu2023-Prompt-Injection-Attacks, jain2023baseline}. 
    \item Retokenization:
Retokenization involves splitting tokens into smaller units using methods like BPEdropout \cite{provilkov2020bpedropout}. This approach can obfuscate malicious instructions or data embedded in the data prompt, neutralizing the adversarial content \cite{jain2023baseline}.
    \item Data Prompt Isolation:
By isolating the data prompt with delimiters such as three single quotes, XML tags, or random sequences, the LM is forced to treat the data prompt strictly as data, ignoring any additional instructions that are not part of the designated task \cite{liu2024promptinjection}.
\end{enumerate}

\subsection{Mitigation for Insecure Output Handling}
Apply the Zero Trust architecture, by treating the LLM as a user whose data most be validated. Validate every output coming from the LLM to the application's backend (classes and functions) before proceesing, displaying to users or saving to the database. \cite{owasp2023}

\section{End-to-End Threat Model for LLM-Powered Apps}

\begin{figure}
    \centering
    \includegraphics[width=1.0\linewidth]{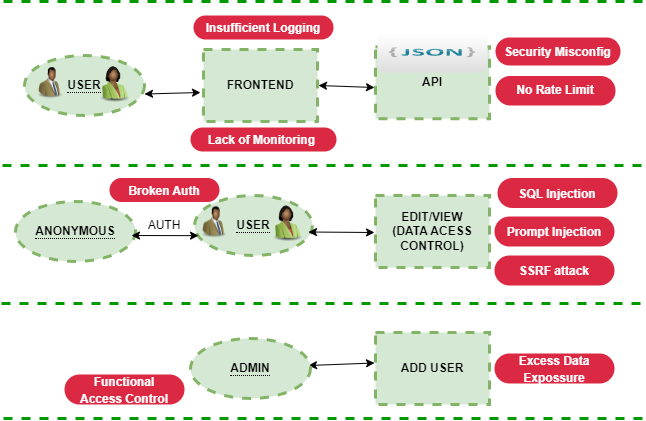}
    \caption{LLM-Application Threat Model}
    \label{fig:enter-label}
\end{figure}

Threat modeling is analyzing representations of a system to highlight concerns about security and privacy characteristics \cite{Threat_Modeling_Manifesto}. Secure software development requires a ‘shift left’ — paying attention to security and privacy early in the life cycle. Threat modelling is a very useful activity for achieving this goal, but for a variety of reasons, organizations struggle to introduce it \cite{securityintelligence_TM_Manifesto}. 

Using one of the best threat modelling framework \cite{Threat_Modeling_Manifesto, machine_generated_tx, tarandachContinuousThreatModeling}, we adapt the Shostack's 4 Question Frame for Threat Modelling \cite{shostack2014threatmodeling, tarandachContinuousThreatModeling}, which presents the following questions:
\begin{enumerate}
    \item What are we working on?
    \item What can go wrong?
    \item What are we going to do about it?
    \item Did we do a good job?
\end{enumerate}

A group of 15 security and privacy researchers \cite{Threat_Modeling_Manifesto}, band together in order to publish a new manifesto designed to guide organizations on their threat modeling journeys, it's called 'Threat Modelling Manifesto' \cite{Threat_Modeling_Manifesto}. Proposed by these researchers are the following values, which are preferred, over others.
We have come to value:
\begin{itemize}
    \item A culture of finding and fixing design issues over checkbox compliance.
    \item People and collaboration over processes, methodologies, and tools.
    \item A journey of understanding over a security or privacy snapshot.
    \item Doing threat modeling over talking about it.
    \item Continuous refinement over a single delivery.
\end{itemize}

The manifesto aims to guide organizations in effective threat modeling, emphasizing its broader utility beyond predicting attacks. It addresses challenges faced by companies in implementing coherent models and highlights the need for accessibility beyond the security team.\cite{johnson_New_Manifesto_CISOs, Threat_Modeling_Manifesto, shostack2014threatmodeling}

The researchers argue that a good threat model goes beyond predicting attacks, aiding in identifying unknown vulnerabilities, estimating risk, and planning security scenarios. The manifesto emphasizes the importance of making threat modeling understandable to a wider audience, including executives and developers, challenging the perception of it as a complex process.\cite{johnson_New_Manifesto_CISOs, Threat_Modeling_Manifesto}

The document encourages organizations to avoid common pitfalls in the threat modeling process and advocates for a workplace culture emphasizing problem-solving, collaboration, and continuous model updates. While not prescribing specific methods, the manifesto outlines high-level values and principles for effective threat modeling, cautioning against overly complex models.

Aligned with agile development principles, the manifesto promotes an iterative and cyclical approach to threat modeling, emphasizing its compatibility with the dynamic nature of security needs. It concludes by suggesting that organizations embracing DevSecOps trends are likely to integrate threat modeling into their product development initiatives. The manifesto aims to facilitate the understanding and adoption of threat modeling practices as a continuous and evolving process. \cite{Threat_Modeling_Manifesto, shostack2014threatmodeling}

\subsection{Threat Modelling with STRIDE \& DREAD }
Creating a threat model involves identifying potential threats and vulnerabilities in a system to understand and mitigate risks. Using STRIDE (Spoofing, Tampering, Repudiation, Information Disclosure, Denial of Service, and Elevation of Privilege) and DREAD (Damage, Reproducibility, Exploitability, Affected Users, and Discoverability) which are two commonly used threat modeling frameworks; we create a threat model for our LLM-powered application:
Threat Model for LLM-Powered Application
\begin{enumerate}
    \item \textbf{Spoofing (STRIDE):}

    Threat: Unauthorized access to the application or API using fake credentials.
    DREAD:
        Damage: High (Potential compromise of sensitive information)
        Reproducibility: High (Could be attempted by malicious actors)
        Exploitability: Medium (Depends on the strength of authentication mechanisms)

    \item \textbf{Tampering (STRIDE):}

    Threat: Modification of input data to the language model for malicious purposes (e.g., prompt injection).
    DREAD:
        Damage: High (Altered or biased model outputs)
        Reproducibility: Medium (Requires understanding of model input mechanisms)
        Exploitability: High (Common attack vector)

    \item \textbf{Repudiation (STRIDE):}

    Threat: Denying the execution of certain actions, such as generating specific outputs from the language model.
    DREAD:
        Damage: Medium (Could lead to disputes or lack of accountability)
        Reproducibility: Medium (Depends on the type of repudiation)
        Exploitability: Low to Medium (May require sophisticated attacks)

    \item \textbf{Information Disclosure (STRIDE):}

    Threat: Unauthorized access to sensitive information generated by the language model.
    DREAD:
        Damage: High (Potential exposure of sensitive data)
        Reproducibility: Medium (Depends on the nature of information disclosure)
        Exploitability: Medium (Depends on security controls)

    \item \textbf{Denial of Service (STRIDE):}

    Threat: Overloading the API with a large number of requests to disrupt normal functioning.
    DREAD:
        Damage: High (Service downtime and impact on user experience)
        Reproducibility: High (Common attack vector)
        Exploitability: High (Potential for abuse).

    \item \textbf{Elevation of Privilege (STRIDE):}

    Threat: Gaining unauthorized access to higher privileges within the application or system.
    DREAD:
        Damage: High (Compromise of critical system components)
        Reproducibility: Low (May require multiple vulnerabilities)
        Exploitability: Low (Depends on the existing security controls)
\end{enumerate}

\subsection*{Threat Modelling with Shostack's 4 Question Frame}

\begin{itemize}
    \item \textbf{What Are We Working On?}
    \begin{itemize}
        \item Development of an LLM-powered application for natural language processing tasks.
    \end{itemize}
    
    \item \textbf{What Can Go Wrong?}
    \begin{itemize}
        \item Identified threats include Spoofing, Tampering, Repudiation, Information Disclosure, Denial of Service, and Elevation of Privilege.
    \end{itemize}
    
    \item \textbf{What Are We Going to Do About It?}
    \begin{itemize}
        \item Implemented multi-factor authentication to address Spoofing.
        \item Applied input validation and content integrity checks to prevent Tampering.
        \item Established robust logging and auditing mechanisms for Repudiation.
        \item Encrypted sensitive data and enforced access controls for Information Disclosure.
        \item Implemented rate limiting and traffic analysis for Denial of Service.
        \item Adopted the principle of least privilege and conducted regular access control reviews for Elevation of Privilege.
    \end{itemize}
    
    \item \textbf{Did We Do a Good Enough Job?}
    \begin{itemize}
        \item Continuous threat modeling, security testing, and feedback mechanisms are in place to ensure ongoing improvement.
    \end{itemize}
\end{itemize}

\begin{table*}[ht]
    \centering
    \caption{Summary of some prompt injection attacks to LLM-powered Applications.}
    \begin{tabular}{|p{4.5cm}|p{6cm}|p{4.5cm}|}
        \hline
        \textbf{Attack Name} & \textbf{Attack Description} & \textbf{Source} \\\hline
        Naive Attack & Concatenate target data, injected instruction, and injected data & Online post: \cite{harang2023, owaspLLM, willisonPromptInjection} \\
        \hline
        Escape Characters & Adding special characters like ``\textbackslash n'' and ``\textbackslash t'' & Arxiv paper: \cite{liu2023prompt} \\
        \hline
        Context Ignoring & Adding context-switching text to mislead the LLM that the context changes & Workshop paper: \cite{perez2022ignore} \\
        & & Arxiv paper: \cite{branch2022evaluating} \\
        & & Online post: \cite{harang2023, willison2023delimiters} \\
        \hline
        Fake Completion & Adding a response to the target task to mislead the LLM that the target task has completed & Online post: \cite{willison2022prompt} \\
        \hline
        Combined Attack & Combining Escape Characters, Context Ignoring, and Fake Completion & Arxiv paper: \cite{liu2023prompt} \\
        \hline
    \end{tabular}
    
    \label{tab:attacks}
\end{table*}

 \begin{table*}[h]
    \centering
    \caption{Threat Ranking for LLM-Powered Application}
    \begin{tabular}{|p{3.9cm}|p{1.9cm}|p{1.9cm}|p{1.9cm}|}
        \hline
        \textbf{Threat} & \textbf{Damage} & \textbf{Reproducibility} & \textbf{Exploitability} \\
        \hline
        Spoofing (STRIDE) & High & High & Medium \\
        \hline
        Tampering (STRIDE) & High & Medium & High \\
        \hline
        Repudiation (STRIDE) & Medium & Medium & Low to Medium \\
        \hline
        Information Disclosure (STRIDE) & High & Medium & Medium \\
        \hline
        Denial of Service (STRIDE) & High & High & High \\
        \hline
        Elevation of Privilege (STRIDE) & High & Low & Low \\
        \hline
    \end{tabular}
    
\end{table*}

\section{Conclusion}
\label{sec:conclusion}

As the technological ecosystem rapidly adopts Large Language Models (LLMs), the need for comprehensive threat modeling and risk analysis becomes imperative to safeguard the applications they empower. This study has extensively examined the threats to LLM-powered applications, highlighting potential attacks such as data poisoning, prompt injection, SQL injection, jailbreaking, and compositional injection attacks. Through an adaptive and holistic threat modeling approach, this paper advocates a robust security framework merging STRIDE and DREAD methodologies, offering concrete mitigation strategies tailored for these sophisticated AI systems.

The exploration of an end-to-end threat model for a custom-built LLM-Doctor \cite{tete_llm_doctor} application validates the framework's feasibility and demonstrates its effectiveness in identifying, categorizing, and mitigating threats. By following a methodical approach, this research provides valuable insights and security paradigms that can be implemented across disciplines and industries where LLMs are utilized. However, the dynamic nature of cyber threats and the continuous evolution of LLMs necessitate an ever-vigilant and responsive approach to threat modeling in AI-powered software development.

\section{Limitations}
This paper focuses on threat modelling and risk analysis for LLM-powered applications, in which the LLM is a third party solution, which is accessed through application programming interface (API) only. Other research may explore the use of in-house LLM.

\section{Recommendations for Further Research}
\label{sec:recommendations}

Based on the findings of this study, several recommendations can be made to guide further research in the area of securing LLM-powered applications:

\begin{enumerate}
    \item \textbf{Dynamic Threat Modeling:} Develop adaptive threat modeling methodologies that evolve with LLM advancements and emerging attack vectors, ensuring continual alignment with the changing threat landscape.
    
    \item \textbf{Automated Response Mechanisms:} Explore the creation of automated systems that can detect and neutralize attacks in real-time, effectively reducing the window of vulnerability for LLM-powered applications.
    
    \item \textbf{Integration with DevSecOps:} Investigate strategies for integrating threat modeling and risk analysis practices within the DevSecOps pipeline to enhance the security posture of applications from the outset.
    
    \item \textbf{Cross-disciplinary Case Studies:} Conduct comprehensive case studies across various industries to validate the applicability of the proposed threat modeling frameworks and to customize defensive strategies accordingly.
    
    \item \textbf{User Education and Awareness:} Assess the impact of user education on mitigating risks posed by social engineering and similar human-targeted attacks in the context of LLM interactions.
 \end{enumerate}

 \bibliographystyle{IEEEtran}
 \bibliography{references}

\vfill

\end{document}